# Performance comparison between *p-i-n* tunneling transistors and conventional MOSFETs

Siyuranga O. Koswatta[*] and Mark S. Lundstrom, *Fellow, IEEE*

School of Electrical and Computer Engineering, Purdue University, West Lafayette, Indiana 47906, USA

Dmitri E. Nikonov, *Senior Member, IEEE*

Technology and Manufacturing Group, Intel Corp., SC1-05, Santa Clara, California 95052, USA

*Abstract* – Field-effect transistors based on band-to-band tunneling (BTBT) have gained a lot of recent interest due to their potential for reducing power dissipation in integrated circuits. In this paper we present a detailed performance comparison between conventional *n-i-n* MOSFET transistors, and BTBT transistors based on the *p-i-n* geometry (*p-i-n* TFET), using semiconducting carbon nanotubes as the model channel material. Quantum transport simulations are performed using the nonequilibrium Green's function formalism including realistic phonon scattering. We find that the TFET can indeed produce subthreshold swings below the conventional MOSFET limit of 60mV/decade at room temperature leading to smaller off-currents and standby power dissipation. Phonon assisted tunneling, however, limits the off-state performance benefits that could have been achieved otherwise. Under on-state conditions the drive current and the intrinsic device delay of the TFET are mainly governed by the tunneling barrier properties. On the other hand, the switching energy for the TFET is observed to be fundamentally smaller than that for the MOSFET, reducing the dynamic power dissipation. Aforementioned reasons make the *p-i-n* geometry well suited for low power applications.

[*] Email address: koswatta@purdue.edu

I. INTRODUCTION

With the continual miniaturization of the MOSFET transistors, power dissipation in integrated circuits has become a major roadblock to performance scaling [1]. For more than 30 years, numerous breakthroughs in device and material design have sustained an exponential increase in system performance [2]. The recent introduction of high-k gate oxides into semiconductor technology has also allowed much needed reduction in gate leakage and improved the scalability of future devices [3]. Nevertheless, the physical operational principles of conventional MOSFETs, based on the thermionic emission of carriers over a channel barrier, have imposed fundamental limits on voltage scaling and the reduction of energy dissipation [2]. The subthreshold swing (*S*) of a conventional MOSFET, which determines the ability to turn off the transistor with the gate voltage ($V_{GS}$), has a fundamental limit of $2.3*(k_B T/q)$ where $k_B$, $T$, and $q$ are the Boltzmann constant, temperature, and the electron charge, respectively (*S* = 60mV/decade at room temperature) [4]. Therefore, the requirement of achieving a large on-state current ($I_{ON}$), while maintaining a small off-state leakage ($I_{OFF}$), has hindered the scaling of the power supply voltage ($V_{DD}$) in recent years [1]. Consequently, a device with *S* below the aforementioned conventional limit is desirable for continued voltage scaling, and thereby reducing power dissipation in circuits.

Field-effect transistors based on the band-to-band tunneling (BTBT) phenomenon are being actively investigated due to their potential for low standby leakage [5-12]. It has been predicted through detailed device simulations that BTBT FETs could produce subthreshold swings below the thermal limit in conventional semiconductor materials such as silicon [13-16], as well as in carbon nanotube (CNT) based transistors [17-21]. Indeed, this has been experimentally demonstrated in CNTs [22-24] and more recently with a silicon based BTBT FET [25]. BTBT occurs in two different transistor geometries; a popular *p-i-n* geometry reported in [5-9, 11, 13, 15-18, 20, 21, 24, 25] (hereafter called the TFET), and the conventional MOSFET geometry used in [12, 19, 22, 23]. In the case of CNT-MOSFETs [22, 23] it has been established that BTBT is dominated by phonon assisted inelastic tunneling that severely deteriorates the device characteristics [19, 26]. On the other hand, phonon scattering has a less dramatic effect on TFETs, and useful device properties are preserved under practical biasing conditions [18, 27].

This paper addresses the important task of a comprehensive comparison of device performance between the *p-i-n* TFET and the conventional *n-i-n* MOSFET geometries. Here, we use CNTs as the model channel material due to many benefits of that system. CNTs allow one-dimensional carrier transport without depletion capacitance effects, and high performance transistors that operate near the ballistic limit have already been demonstrated [28-30]. They also have a direct energy bandgap and small carrier effective masses that are favorable for BTBT devices [31]. Furthermore, a detailed simulation framework has been developed for modeling carrier transport through CNT transistors [32-35], and benchmarked against experiments [26, 29, 35]. Therefore, many realistic aspects, such as the effect of phonon scattering on device performance, have been comprehensively explored in the case of CNT based MOSFETs [19, 26, 36] as well as TFETs [27]. Previous work has also compared CNT transistor performance to that of silicon transistors [37-39] and to that based on silicon nanowires [40]. Here we use similar device metrics to compare the performance between TFETs and MOSFETs using a uniform simulation environment for both the devices.

The organization of the rest of the paper is as follows. For the sake of clarity of the ensuing discussion, section II presents a brief description of the simulation procedure used in this study. Section III compares the off-state characteristics for the two geometries. The on-state performance comparison is presented in section IV where we observe an important difference in switching energy related to their switching capacitances. Therefore, section V provides insights into the origin of device capacitance at the quantum capacitance limit. Finally, section VI presents the conclusions.

## II. METHOD

The model device structure used in this study, shown in Fig. 1, has a cylindrical wrap-around gate and doped source/drain regions. We use the following device parameters and $T = 300K$ unless specified otherwise. A (13,0) zigzag CNT with intrinsic channel length, $L_{ch}$ = 15nm, doped source/drain regions with $L_{S,D}$ = 20nm, high-k HfO$_2$ oxide ($k$ = 16) with $t_{OX}$ = 2nm have been used. The source/drain linear doping concentration is 0.8/nm which can be compared with the carbon atom density for a (13,0) CNT of 122/nm. When comparing the two device geometries, the source region is doped either *p*- or *n*- type accordingly, keeping all other parameters identical. It has been observed that having high-k gate oxide over the doped source region increases the source-channel fringing fields, which results in smaller tunneling currents in the case of TFETs. Therefore, in this study we have removed the high-k oxide from the source/drain regions as shown in Fig. 1 (in realistic device fabrication these regions could be filled with a low-k spacer dielectric).

We have performed dissipative quantum transport calculations using the non-equilibrium Green's function (NEGF) formalism [41]. A self-consistent electrostatics solution is obtained by solving the 2D-Poisson's equation (in the $\hat{r}$ and $\hat{z}$ directions) using the finite difference scheme [33]. A detailed description of the simulation procedure is given in [33]. We summarize the main equations here for the sake of clarity. The device Green's function, $G$, at an energy $E$ in the presence of electron-phonon (e-ph) scattering is given by [41], $G = \left[ EI - H_{pz} - \Sigma_S - \Sigma_D - \Sigma_{scat} \right]^{-1}$ where, $I$ is the identity matrix, and $H_{pz}$ is the device Hamiltonian matrix in the nearest-neighbor $p_z$ tight-binding basis [31, 33]. Here, the mode-space treatment for carrier transport is used [33, 42], where we consider the lowest conduction band and the highest valence band with two-fold spin and two-fold valley degeneracies [31]. The self-energy functions, $\Sigma_{S,D}$ and $\Sigma_{scat}$, arise due to coupling to the semi-infinite source/drain contacts and due to e-ph interaction, respectively. The energy dependence and the matrix representation of these functions are implicit [33]. Level-broadening due to contact coupling is then given by [41], $\Gamma_{S,D} = i\left( \Sigma_{S,D} - \Sigma_{S,D}^{\dagger} \right)$, where $\Sigma^{\dagger}$ is the conjugate transpose of the self-energy matrix. Under ballistic conditions the spectral function can be separated into its source and drain contributions, respectively [41];

$$A_{S,D} = G\Gamma_{S,D}G^{\dagger} \tag{1}$$

where the diagonal elements of $A_{S,D}$ are related to the local density of states (LDOS$_{S,D}$) evolving from the corresponding contact [41].

The in/out-scattering functions that account for coupling to the source/drain reservoirs is given by

$$\Sigma_{S,D}^{in} = \Gamma_{S,D}f_{S,D} \ , \ \Sigma_{S,D}^{out} = \Gamma_{S,D}\left(1 - f_{S,D}\right), \tag{2}$$

where $f_{S,D}$ are the contact Fermi distributions. The in/out-scattering functions for e-ph interaction of an optical phonon (OP) mode with energy $\hbar\omega$ are given by [41]

$$\Sigma_{scat}^{in,out}(E) = D_0\left(n_\omega + 1\right)G^{n,p}(E \pm \hbar\omega) + D_0 n_\omega G^{n,p}(E \mp \hbar\omega), \tag{3}$$

where $D_0$ is the e-ph coupling parameter calculated according to [43]. The electron/hole correlation functions, $G^{n,p}$, are given by [41]

$$G^{n,p} = G\left[\Sigma_S^{in,out} + \Sigma_D^{in,out} + \Sigma_{scat}^{in,out}\right]G^{\dagger}. \tag{4}$$

From eqs. (4), (1), and (2), it is seen that, under ballistic conditions (i.e. $\Sigma_{scat}^{in,out} = 0$), the electron/hole distribution throughout the device is determined by the occupation of the respective local density of states, LDOS$_{S,D}$, by the corresponding reservoir Fermi functions, $f_{S,D}$.

In treating e-ph scattering (eq. (3)) we are assuming the scattering functions to be diagonal due to the local interaction approximation [33]. OP scattering by 190meV longitudinal optical (LO) mode, 180meV zone-boundary (ZBO) mode, 26meV radial-breathing mode (RBM), and acoustic phonon (AP) scattering by the longitudinal acoustic (LA) mode have been considered in the case of the (13,0) CNT [33]. The phonon population in eq. (3) is assumed to be in equilibrium with the external thermal bath, with the number, $n_\omega$, given by the Bose-Einstein distribution,

$$n_\omega = \left(\exp(\hbar\omega/k_B T) - 1\right)^{-1}. \quad (5)$$

The phonon emission mediated processes are described by the first term in the right hand side of eq. (3); the second term corresponds to that of phonon absorption. Finally, the current through the device from site $z$ to $(z+1)$ in the nearest-neighbor tight-binding scheme is given by [33, 44]

$$I_{z \to z+1} = \frac{4ie}{\hbar} \int_{-\infty}^{+\infty} \frac{dE}{2\pi} \left[ H_{pz}(z, z+1) G^n(z+1, z) - H_{pz}(z+1, z) G^n(z, z+1) \right], \quad (6)$$

where the lower and upper diagonal elements of the Hamiltonian and the electron correlation function have been used. Under ballistic conditions, eq. (6) further simplifies to the Landauer equation [41],

$$I = \frac{4e}{\hbar} \int_{-\infty}^{+\infty} \frac{dE}{2\pi} T(E)(f_S - f_D), \quad (7)$$

with the transmission coefficient given by, $T(E) = Trace\left[\Gamma_S G \Gamma_D G^\dagger\right]$. The efficient numerical algorithms of [44] have been employed in our computational simulations.

III. COMPARISON OF THE OFF-STATE OPERATION

A. *Subthreshold slope, Off-current ($I_{OFF}$), and Standby power dissipation ($P_{standby}$)*

One of the main attractions for BTBT transistors is their potential to reduce off-state leakage, and in turn, standby power dissipation ($P_{standby}$) in circuits. This is achieved through subthreshold operation with $S$ below the conventional limit in these devices.

Figure 2 compares the temperature dependence of the transfer characteristics ($I_{DS}$-$V_{GS}$) for the two geometries. The ballistic results (solid curves) are discussed first. In Fig. 2(a) it is observed that we obtain ideal subthreshold operation with $S = 60$mV/dec (at $T = 300$K) due to the superior electrostatic control by the wrap-around gate. At higher temperatures, however, $S$ degrades proportionately. This can be easily understood by observing Fig. 3(a) showing the thermionic emission mechanism in the off-state of a conventional MOSFET. The high energy tail of the Fermi distribution grows with temperature as $\sim \exp(-E/k_B T)$ leading to the aforementioned degradation of $S$. Furthermore, in integrated circuits this results in higher off-state leakage currents and $P_{standby}$. This could lead to a positive feedback mechanism between the two, known as thermal runaway, that could ultimately destroy the circuit [45].

On the other hand, the ballistic results for the TFET (Fig. 2(b)) clearly shows $S < 60$mV/dec operation at room temperature. This is easily understood by examining Fig. 3(b) where the high energy tail of the Fermi distribution for electrons lies inside the *p*-type source bandgap region. Therefore, when the conduction band in the channel is pulled above the valence band of the source, an abrupt reduction in device current is expected, which leads to $S$ values much smaller than the conventional limit [21]. Figure 2(b) provides an interesting observation that the off-state current under ballistic transport does not significantly degrade at elevated temperatures. This is due to the elimination of high energy thermal injection within the source bandgap region. There is a slight increase in subthreshold current at higher temperatures related to the broadening of the Fermi distribution near $E_{FS}$ (see Fig. 3(b)). The possibility of achieving off-stage leakage currents that do not degrade at higher temperatures is an attractive feature of TFETs that could potentially alleviate the thermal runaway problem mentioned earlier.

The relative benefits of the TFET over the MOSFET geometry in the off-state can be better compared through the $I_{OFF}$ vs. $I_{ON}$ (at a constant $V_{DD}$) results shown in Fig. 4(a) for ballistic operation. Here, the $I_{OFF}$-$I_{ON}$ curves are generated by scanning the $I_{DS}$-$V_{GS}$ results in Fig. 2 with a constant $V_{DD}$ bias window as explained in Ref. [37]. In Fig. 4(a) the increase in $I_{OFF}$ at smaller $I_{ON}$ values ($I_{ON} \leq 0.6\mu A$/tube) observed for the TFET is due to ambipolar conduction seen in Fig. 2(b). Figure 4(a) clearly shows the suppression of $I_{OFF}$ degradation at higher temperatures under ballistic transport in the case of TFETs compared to MOSFETs. Furthermore, the shaded region of Fig. 4(a) corresponds to the range of device biasing conditions where the TFET outperforms the MOSFET. Within this region it is observed that the former has a smaller $I_{OFF}$ (thus smaller $P_{standby}$) at a given $I_{ON}$ (looking vertically). Conversely, the TFET can deliver a larger $I_{ON}$ at a given $I_{OFF}$ (looking horizontally). It is noted that in this region the TFET can only deliver a few μA of drive current per CNT. Therefore, these devices might be better suited for low power applications with moderate drive current requirements.

### B. *Influence of phonon scattering*

The influence of phonon scattering on the transfer characteristics is shown by the dashed curves of Fig. 2 (see Refs. [19, 26, 27, 36] for detailed information on each geometry). In Fig. 2(a) it is observed that phonon scattering has only a small effect on the subthreshold properties of the MOSFET which are dominated by the thermionic emission component of current conduction. Nevertheless, at small gate biases the onset of ambipolar conduction is seen due to phonon assisted inelastic tunneling that turns on BTBT in CNT-MOSFETs at a larger voltage [19, 26]. In the case of the TFET in Fig. 2(b), it is observed that $S$ degrades in the presence of phonon scattering even though $S <$ 60mV/dec is still attained. This deterioration is due to phonon absorption assisted transport playing an important role under off-state biasing conditions (see Fig. 3(b)) [27]. More importantly, the subthreshold current becomes temperature dependent due to larger phonon occupation (eq. (5)) at higher temperatures that increases phonon absorption assisted transport (second term of eq. (3)).

Examining Fig. 4(b) we observe that there still exists a possible biasing region (shaded) where the TFET outperforms the MOSFET geometry. The significant increase in $I_{OFF}$ at higher temperatures is clearly observed for the TFET. Off-state leakage that is about an order of magnitude smaller compared to the MOSFET can still be attained at both room and elevated temperatures. In comparing Figs. 4(a) and 4(b), however, it is clear that the ballistic operation of TFETs could have reduced the off-state leakage significantly, especially at higher temperatures. Therefore, it can be summarized that the TFET can indeed deliver superior subthreshold characteristics under realistic transport conditions, but phonon scattering deteriorates the beneficial features that could have been attained otherwise.

C.    *Drain induced off-state degradation*

In this section we examine the effect of the drain bias on subthreshold properties. In a conventional MOSFET this could lead to the well known drain induced barrier lowering (DIBL) effect [4]. In the case of a TFET in the off-state, as shown in Fig. 3(b), it is not clear whether DIBL would have a similar effect since the high energy tail of the Fermi distribution is already within the source bandgap region. Therefore, in order to study the effect of the drain bias on subthreshold current we use a slightly modified device geometry compared to Fig. 1 that allows drain field penetration into the channel region in our wrap-around gate structure. In this section we use a (10,0) CNT with $t_{OX}$ = 5nm and $SiO_2$ ($k$ = 3.9) for the gate oxide that covers the full length of the tube including the source/drain regions. All other device parameters are similar to the previous case, and ballistic transport simulations are performed.

The transfer characteristics and their drain bias dependence for the two geometries are compared in Fig. 5. We observe that the CNT-MOSFET still retains the well tempered operation with very small DIBL. On the other hand, the TFET shows a significant bias dependence similar to the DIBL effect of a MOSFET. A closer examination of the energy bands (Fig. 6), however, provides insight into the origin of this bias dependence. First of all, in the case of the MOSFET in Fig. 6(a) it is observed that the top of the channel barrier does not get pulled down significantly at larger $V_{DS}$; thus, the smaller DIBL seen for this case. On the other hand, in Fig. 6(b) there is a significant shortening of the channel barrier width at large $V_{DS}$ for the case of the TFET. The transmission coefficient for direct electron tunneling through the channel region increases exponentially with decreasing barrier width. Therefore, as expected from eq. (7) the off-state current increases significantly. This effect for TFETs observed here can be identified as drain induced barrier shortening (DIBS). DIBS could be important for a highly scaled device where short channel effects are considerable. If the original channel barrier width (~ $L_{ch}$) were long enough, the actual magnitude of direct tunneling current would be very small even in the presence of barrier shortening effects.

## IV. COMPARISON OF THE ON-STATE PERFORMANCE

### A. On-current ($I_{ON}$)

One of the main concerns for BTBT based transistors has been their ability to deliver adequate drive currents. The use of TFETs with only one tunneling barrier for carriers as opposed to BTBT in MOSFETs [12, 19, 22, 23] where there are two, the on-current for the former has improved. And, even though there have been many optimization strategies proposed for TFETs in order to improve $I_{ON}$ further [13, 16, 18, 20, 21] it still remains a challenge. Figure 7 compares $I_{DS}$-$V_{GS}$ in linear scale for the two devices shown in Fig. 1. It is observed that the drive current for the TFET is about 3x smaller than that for the MOSFET. If the high-k oxide covers the CNT throughout, including the source region, $I_{ON}$ further degrades by about 18x compared to the MOSFET (not shown). In Fig. 7, however, it is observed that phonon scattering has only a minor effect on TFET on-state current (reduces by ~ 10%) compared to the MOSFET (reduces by ~ 16%). This is because in the case of the former the back-scattered carriers in the channel region have a larger probability of being reflected back by the source-channel tunneling barrier, ultimately escaping into the drain. Thus, DC current transport is not significantly affected by phonon scattering. Therefore, from Fig. 7 it can be noted that the tunneling barrier properties of a TFET have a more dominant effect on the drive current, and the channel mobility itself has only a comparatively minor influence [13, 15, 21, 27].

*B. Intrinsic device delay metric (τ)*

Intrinsic device delay (τ) is an important performance metric that corresponds to intrinsic limitations on switching speed and AC operation of a transistor [4]. In this work the switching speed is calculated as $\tau = (Q_{ON} - Q_{OFF})/I_{ON}$, instead of the traditional equation, $\tau = C_g V_{DD}/I_{ON}$, [4] due to the strong bias dependence of gate capacitance, $C_g$ (see Section V). Here, $Q_{ON,OFF}$ is the total charge induced in the transistor in the on- and off-states, respectively (calculated similar to $I_{ON,OFF}$ with a constant $V_{DD}$ bias window [37]). Thus, τ accounts for any additional charging induced by fringe capacitance effects. Figure 8 shows the τ vs. $I_{ON}/I_{OFF}$ comparison for the TFET and the MOSFET. Surprisingly, we observe that τ for the former is comparable to that of the latter even though the MOSFET has a much larger drive current (Fig. 7). At larger $I_{ON}/I_{OFF}$ ratios (> $10^4$) the TFET is even faster. The main reason for this behavior is the amount of charge involved in the on-off transition of a TFET is considerably smaller compared to that for the MOSFET (see Section V). Device delay, however, increases significantly (not shown) when high-k oxide covers the full length of the CNT including the source region due to the reduction in $I_{ON}$.

In Fig. 8 it is also observed that phonon scattering increases τ for both devices. Even though the drive current for the TFET does not deteriorate substantially in the presence of phonon scattering (Fig. 7) the degradation of τ is comparatively larger. This is due to the occupation of negative going states (-k) in the channel in the presence of back-scattering, and the occupation of low energy states with smaller band velocities, which increase the average transit time for carriers. A similar effect has also been reported in Schottky barrier CNTFETs [46]. Finally, even though the intrinsic delay in Fig. 8 is comparable for the two geometries, the TFET could become significantly slower in the presence of a load capacitance (such as a long interconnect). In such cases the actual drive current of the device becomes important and the MOSFET would have a considerable advantage (Fig. 7).

C.  *Power-delay product (PDP) and Dynamic power dissipation ($P_{dynamic}$)*

Power-delay product (PDP) is the switching energy required for on-off transition of a transistor. It is a measure of the dynamic power dissipation, $P_{dynamic} = \alpha(PDP)f$, where $f$ is the operating frequency and $\alpha$ the activity factor [4]. In this work we calculate PDP by, $PDP = (Q_{ON} - Q_{OFF})V_{DD}$, that corresponds to charging of the MOS capacitor under the voltage bias $V_{DD}$. Figure 9 compares the PDP vs. $I_{ON}/I_{OFF}$ for the two geometries. Here, it is observed that the TFET has a smaller switching energy compared to the MOSFET. Furthermore, the relative shapes of the two curves appear to be fundamentally different; the MOSFET curve is concave downwards while that for the TFET is concave upwards, thus resulting in a smaller PDP under practical biasing conditions (similar distinctions for the two geometries have been observed under various device parameters [18]).

These apparent fundamental differences for the two can be attributed to their total gate capacitances, $C_g$ (see dashed lines of Fig. 10). Here, $C_g = dQ_{tot}/dV_{GS}$ where $Q_{tot}$ is the total charge induced throughout the transistor. In Fig. 10 we observe that the $C_g$-$V_{GS}$ curves for the two geometries at finite $V_{DS}$ have very different shapes (see Section V for details). Now, note that the switching energy of a transistor can be written in an alternative form, $PDP \approx C_{ave}V_{DD}^2$ [4] where the capacitance $C_{ave}$ is an average value determined from the $C_g$-$V_{GS}$ curve for the appropriate gate bias range, from 0 to $V_{DD}$ (= 0.3V in this case). In Fig. 9, going from smaller to larger $I_{ON}/I_{OFF}$ ratios (i.e. left to right) corresponds to moving from the on-state to the off-state in the $C_g$-$V_{GS}$ curves of Fig. 10 (i.e. right to left). Therefore, $C_{ave}$ will also change accordingly. It is easily seen that the integral under the dashed curve for TFET (Fig. 10(b)) is always smaller than that for the MOSFET (Fig. 10(a)).

Thus, it is apparent that the observed differences in the shapes of the PDP curves (Fig. 9) for the two geometries are related to their non-equilibrium device capacitances (Fig. 10). Furthermore, because of the use of ultra thin high-k gate oxides, the devices operate in the quantum capacitance limit. Therefore, it is evident that the important differences observed for the switching energy of the two devices are in fact related to their quantum capacitances. At this point, we take a closer look at the origin of these distinctions for the two geometries since that provides useful insights into the fundamental differences in the relevant device physics.

## V. DEVICE OPERATION AT THE QUANTUM CAPACITANCE LIMIT

The continual increase in gate oxide capacitances (as the device sizes scale down), along with the importance of quantum confined structures, have made the quantum capacitance limit of device operation increasingly relevant; i.e. $C_{OX} \gg C_Q$ condition where $C_{OX}$ and $C_Q$ are the gate oxide and quantum capacitance, respectively [47, 48]. And, $C_Q$ is related to the average density of states (DOS) near the Fermi level, $C_Q \sim DOS(E_F)$ [41]. As discussed below, the DOS of TFET structures can become significantly small, thus the aforementioned condition can be easily achieved. In this case the gate capacitance is dominated by $C_Q$ itself; $C_g = (C_{OX} C_Q)/(C_{OX} + C_Q) \approx C_Q$.

Figure 10 compares the bias dependence of $C_g$ for the two devices under dissipative transport (similar behavior is obtained under ballistic transport as well). It is observed that the equilibrium $C_g$-$V_{GS}$ curves (solid) for the two are very similar, and carry the characteristic signature of the carbon nanotube DOS [31]. This relationship has also been experimentally verified in the case of CNT-MOSFETs [49]. At larger $V_{DS}$, however, interesting differences arise. First of all, in the case of the MOSFET the initial peak splits into two. This is because the negative going (-$k$) half of channel DOS is filled only at a larger $V_{GS}$ compared to the positive going (+$k$) ones [48]. On the other hand, $C_g$-$V_{GS}$ curve for the TFET remains notably small up to larger gate biases. This means that the charge induced in the channel for the TFET is considerably lower compared to that for the MOSFET. This difference can be clearly observed in the energy-position resolved electron distribution function (eq. (4)) shown in Fig. 11. There, the +$k$ states in the channel of the MOSFET that are below the source Fermi energy ($E_{FS}$) are well occupied. On the other hand, in the case of the TFET even though the conduction band in the channel is well below the source $E_{FS}$ the channel states are relatively empty. This is due to the presence of the tunneling barrier that hinders carrier injection into the channel from the source reservoir [21, 48].

Since $C_Q$ in fact originates from filling of the channel states by the source and drain reservoirs, it is instructive to distinguish these states by which of the two contacts of the TFET they would be filled. This is achieved by looking at the contact resolved LDOS shown in Fig. 12 (from eq. (1)), which, we stress, can be strictly separated only in the ballistic approximation. In Fig. 12(a) one observes that there is only a small amount of source-evolving states inside the channel which are filled by that reservoir. Therefore, they have only a small contribution to $C_Q$. On the other hand, there is a large number of drain-evolving states inside the channel; both $-k$ states, as well as $+k$ states that originate from the reflection of the former against the tunneling barrier (Fig. 12(b)). These states are, however, not filled by the drain Fermi reservoir at large $V_{DS}$. They get filled only at larger gate biases, and subsequently increase $C_Q$ as observed by the dashed curve of Fig. 10(b). On the other hand, at small drain biases these states are easily filled, and dominate $C_Q$ (solid curve of Fig. 10(b)). Note that TFET would have very small $C_Q$ under non-equilibrium conditions, and would easily get into the quantum capacitance limit of operation [21]. Furthermore, the characteristic differences in drain bias dependence of $C_Q$ for the MOSFET and the TFET should be readily distinguishable from an experiment similar to [49].

## VI. CONCLUSION

This paper presented a comprehensive comparison of device performance between the conventional MOSFET and the *p-i-n* TFET geometries. It was confirmed that the TFET can indeed operate with a subthreshold swing below the 60mV/decade conventional limit, thereby reducing off-state leakage and standby power dissipation. Phonon assisted tunneling tends to deteriorate the desirable subthreshold characteristics of a TFET that could have been achieved under ballistic conditions. Under on-state operation, the drive current and the switching speed of a TFET are dominated by the tunneling barrier properties, and phonon scattering comparatively has only a minor effect. On the other hand, at the quantum capacitance limit of device operation, the switching energy of a TFET is observed to be fundamentally smaller compared to that of a MOSFET. Therefore, the *p-i-n* TFET geometry is expected to be a strong candidate for future low power applications.

*Acknowledgment* – Authors thank Prof. Joerg Appenzeller of Purdue University for many stimulating discussions on tunnel transistors. S.O.K thanks the Intel Foundation for Ph.D. Fellowship support. Computational support was provided by the NSF Network for Computational Nanotechnology (NCN).

List of Figure Captions.

Fig 1. The modeled device geometry used in this study with cylindrically symmetric wrap-around gate electrode (see text for device parameters). The high-k oxide is removed from source/drain regions in order to reduce the fringing fields that adversely affect the drive current for the *p-i-n* TFET.

Fig 2. $I_{DS}$-$V_{GS}$ dependence on temperature for, (a) *n-i-n* MOSFET and, (b) *p-i-n* TFET under ballistic and dissipative transport. The latter has reduced temperature dependence under ballistic conditions. Phonon assisted tunneling can, however, degrade the subthreshold characteristics.

Fig 3. Band diagram and the source Fermi distribution for, (a) *n-i-n* MOSFET and, (b) *p-i-n* TFET. In the latter, high-energy part of the source distribution is cutoff by the bandgap region leading to < 60mV/decade subthreshold swing. Phonon assisted tunneling becomes important under these conditions.

Fig 4. $I_{OFF}$ vs. $I_{ON}$ dependence on temperature at $V_{DD} = 0.3V$ under, (a) ballistic and, (b) dissipative transport. Shaded region is where the *p-i-n* TFET has an advantage over the *n-i-n* MOSFET due to larger $I_{ON}$ with a smaller $I_{OFF}$. Temperature dependence of $I_{OFF}$ for the *p-i-n* TFET is also smaller than that for the latter.

Fig 5. Dependence of subthreshold properties on the drain bias for, (a) *n-i-n* MOSFET and, (b) *p-i-n* TFET under ballistic transport (the device geometry is slightly modified from that for the rest of the paper. See text for details). The *n-i-n* MOSFET shows small DIBL compared to the *p-i-n* TFET. For the latter the off-current is increased at high $V_{DS}$ due to drain induced barrier shortening (DIBS) (see Fig 6).

Fig 6. Band diagram in the off-state for, (a) *n-i-n* MOSFET and, (b) *p-i-n* TFET at different drain biases. For the latter, drain induced barrier shortening (DIBS) is observed which increases the tunneling current through the channel barrier exponentially.

Fig 7. Linear $I_{DS}$-$V_{GS}$ comparison for the *n-i-n* MOSFET and *p-i-n* TFET under ballistic and dissipative transport. The on-current for the latter is reduced due to the presence of the tunneling barrier.

Fig 8. Intrinsic device delay metric (τ) vs. $I_{ON}/I_{OFF}$ comparison. Surprisingly, *p-i-n* TFET shows similar delay compared to the *n-i-n* MOSFET even though the former has a smaller drive current (Fig. 7). Also, *p-i-n* TFET even becomes faster at larger $I_{ON}/I_{OFF}$ operating regime. In the presence of a load capacitance, however, the actual drive current will become important and the *p-i-n* TFET could be relatively slower.

Fig 9. Comparison of the power-delay product (PDP = switching energy). *p-i-n* TFET has a significant benefit here, and shows a fundamentally different behavior compared to the *n-i-n* MOSFET.

Fig 10. Total device capacitance ($C_g$) vs. $V_{GS}$ calculated from $dQ_{tot}/V_{GS}$ for, (a) *n-i-n* MOSFET and, (b) *p-i-n* TFET under dissipative transport. At small $V_{DS}$ both devices show similar characteristics. However, at larger $V_{DS}$ a fundamentally different behavior is observed; for the *p-i-n* TFET device capacitance remains small until larger gate biases are applied.

Fig 11. Energy-position resolved electron distribution for, (a) *n-i-n* MOSFET and, (b) *p-i-n* TFET under ballistic transport at $V_{GS} = 0.5$V, $V_{DS} = 0.3$V. A significantly higher occupation of channel states is observed for the former.

Fig 12. Reservoir resolved LDOS for the *p-i-n* MOSFET at $V_{GS} = 0.5$V, $V_{DS} = 0.3$V: (a) source-evolving states ($LODS_S$), (b) drain-evolving states ($LDOS_D$). There is a small amount of source-evolving states inside the channel due to the presence of the tunneling barrier. These are the states filled by the source Fermi distribution. Interestingly, there is a significant amount of drain-evolving states inside the channel but they are not filled at large $V_{DS}$.

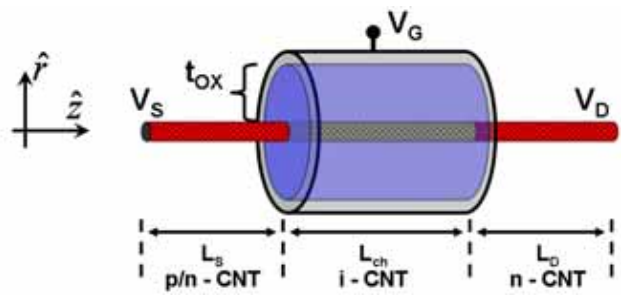

Figure 1

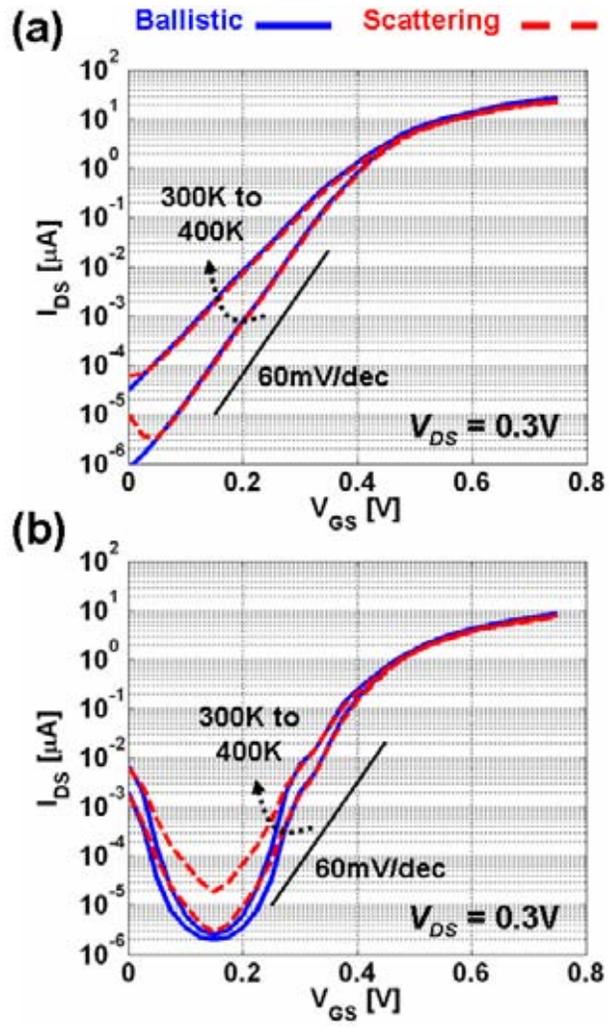

Figure 2

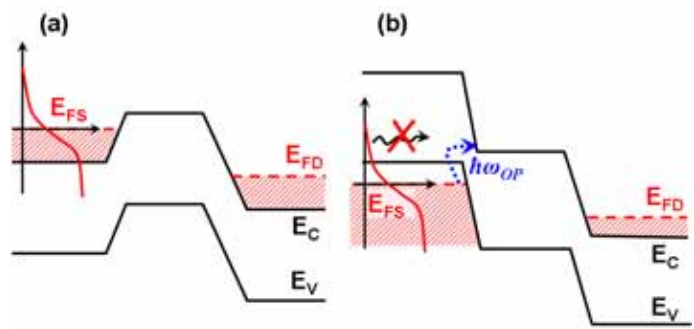

Figure 3

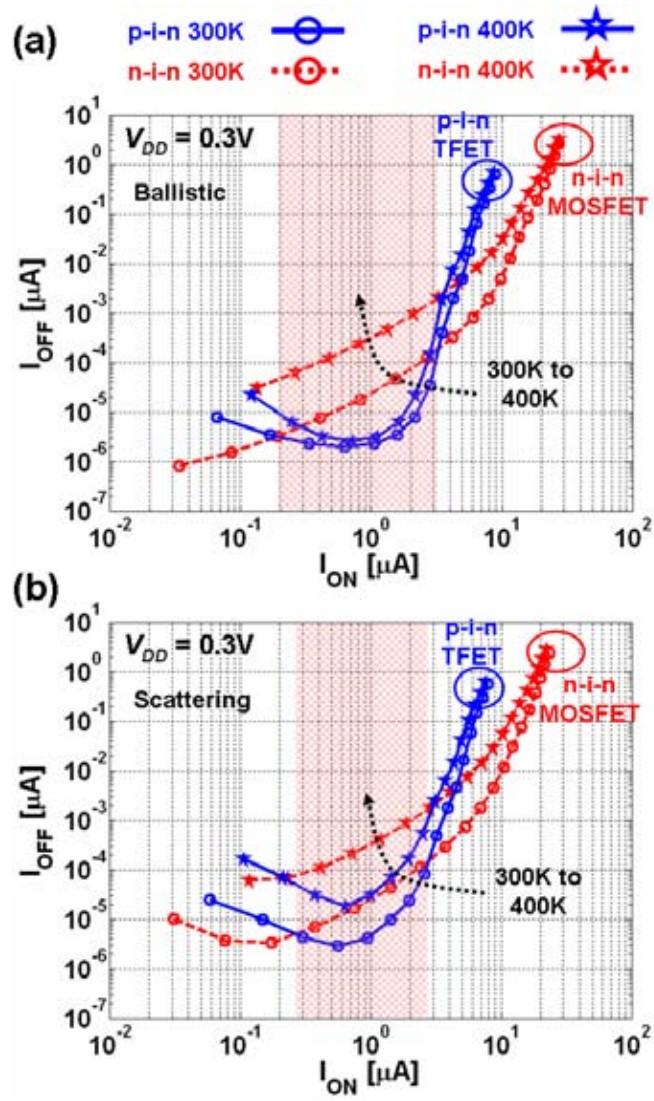

Figure 4

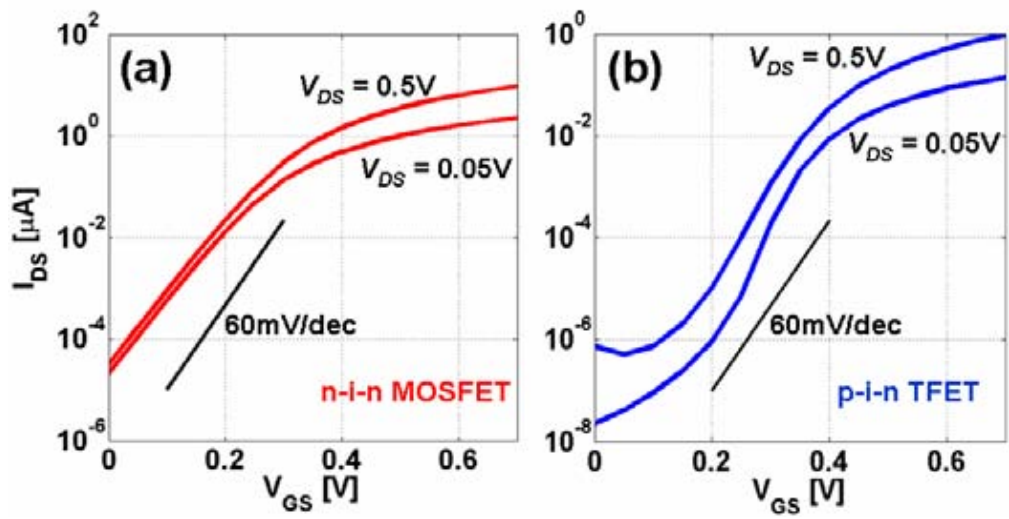

Figure 5

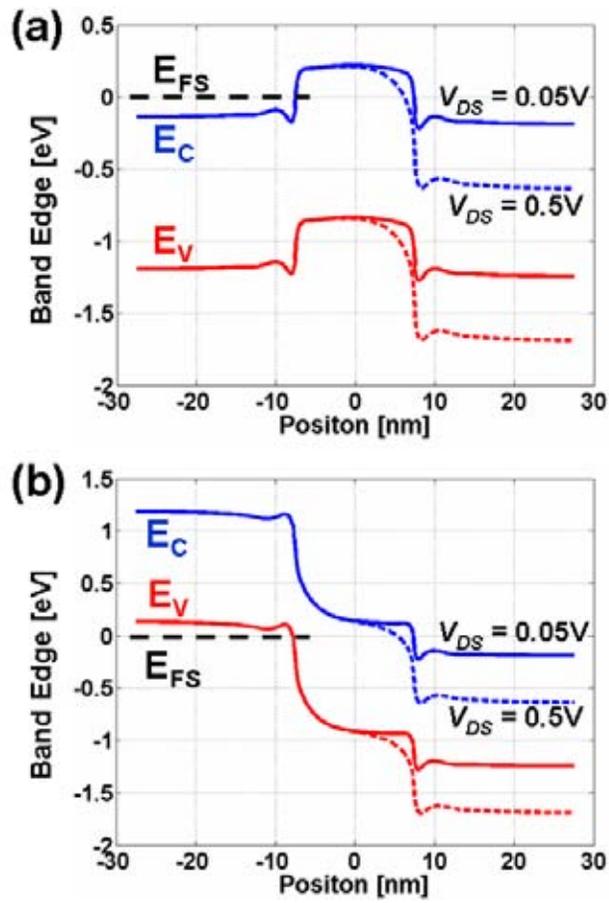

Figure 6

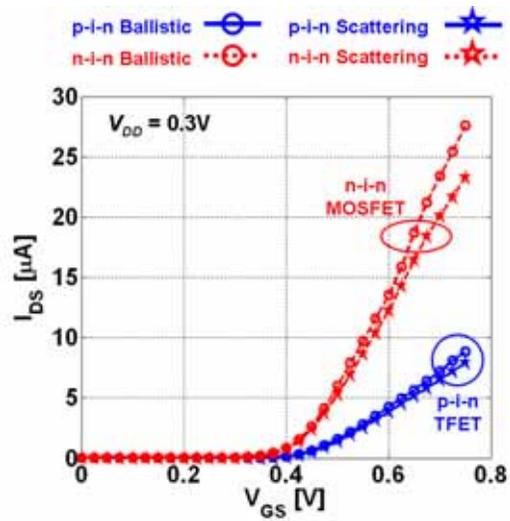

Figure 7

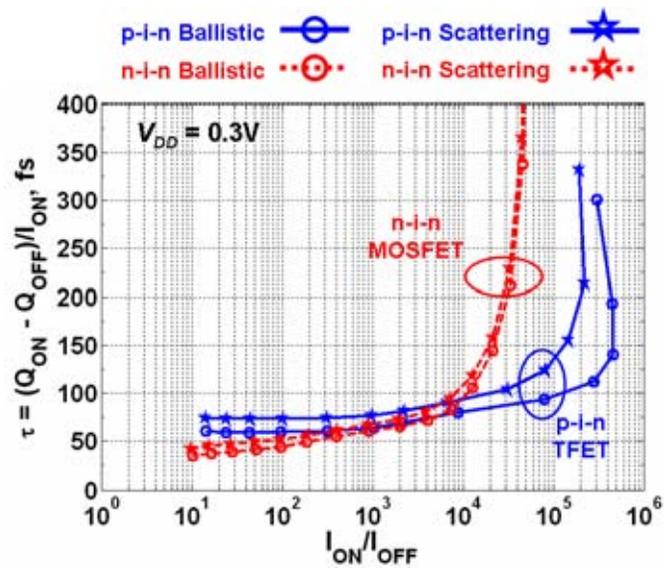

Figure 8

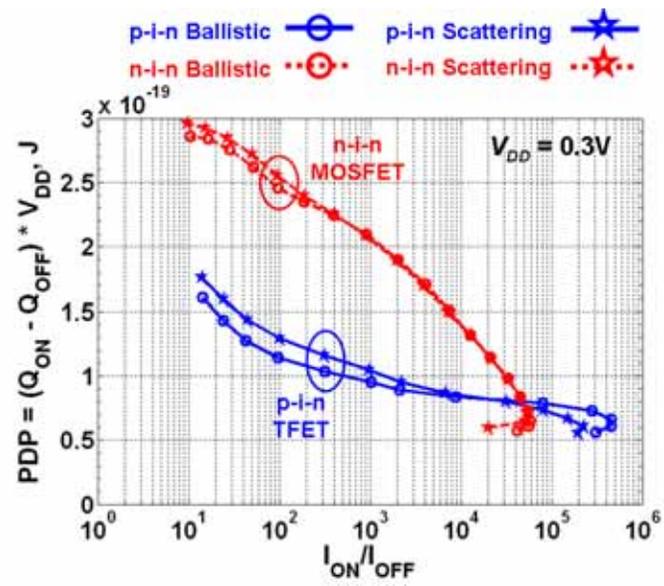

Figure 9

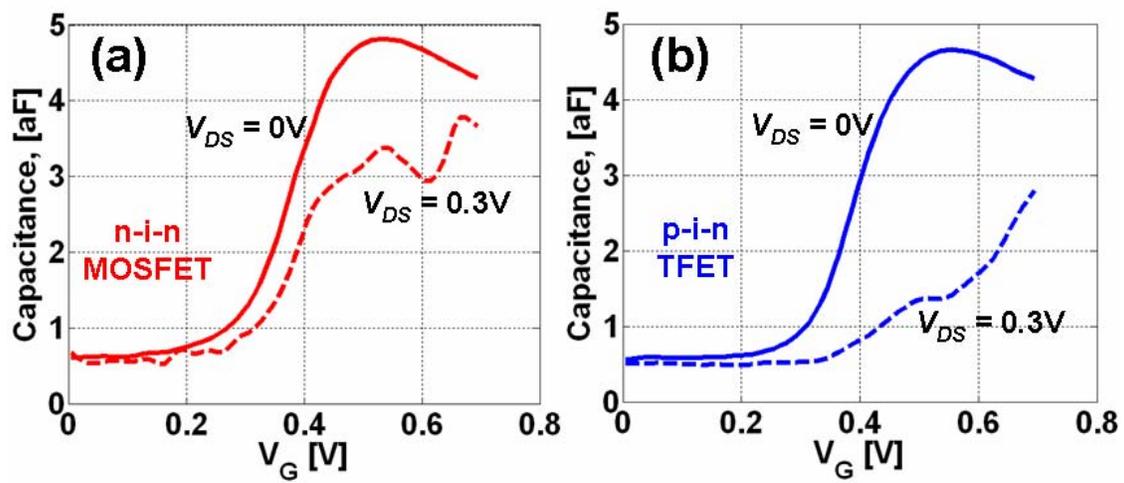

Figure 10

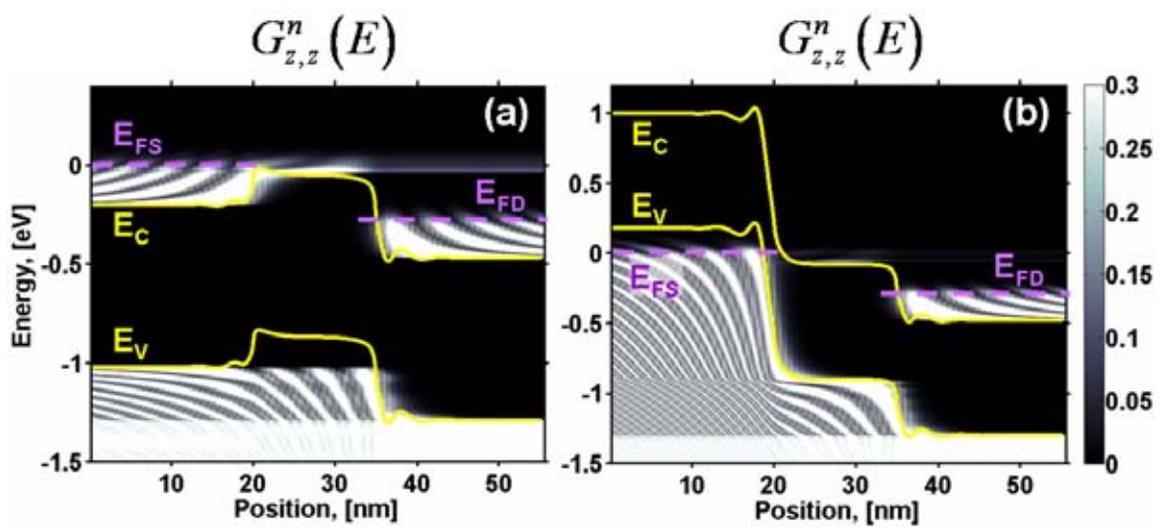

Figure 11

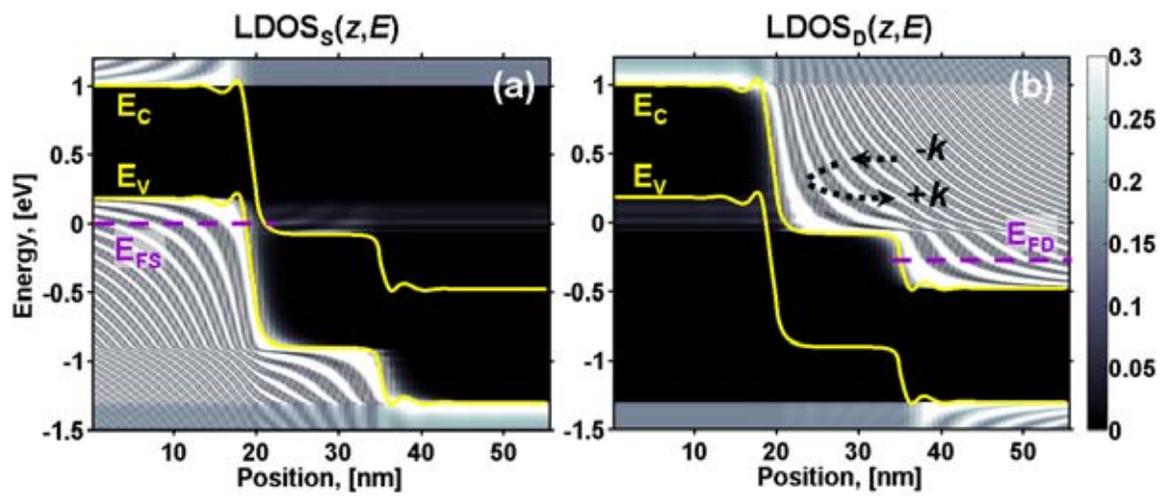

Figure 12